\documentstyle[11pt,newpasp,twoside,psfig]{article}

\index{summary}
\index{instructions}
\index{template}

\def\edcomment#1{\iffalse\marginpar{\raggedright\sl#1\/}\else\relax\fi}
\reversemarginpar

\begin{document}
\title{The Vela pulsar wind nebula at 6cm}
\author{Dion Lewis$^{1,2}$, Richard Dodson$^1$, David McConnell$^2$ and Avinash Deshpande$^3$}
\affil{$^1$University of Tasmania, Department of Mathematics and Physics, GPO Box 252-21, Hobart Tasmania 7001 AUSTRALIA}
\affil{$^2$Australia Telescope National Facility, CSIRO, PO Box 76, Epping NSW 1710 AUSTRALIA}
\affil{$^3$Raman Research Institute, C.V. Raman Avenue, Sadashivanagar P.O. Bangalore 560080 INDIA}

\begin{abstract}  Observations using the Australia Telescope Compact Array at a wavelength of 6~cm have uncovered the radio counterpart to the compact X-ray nebula surrounding the Vela pulsar.  Two lobes were found oriented about the spin axis of the pulsar, starting at the edge of X-ray emission, they extend to three times the size.  The northern lobe has a bright, defined edge and an integrated flux of 0.14 Jy, while the southern lobe of 0.12 Jy is more diffuse.
\end{abstract}

\section{Introduction}  High resolution images from Chandra X-ray observations of the Vela pulsar have shown (Helfand, Gotthelf, \& Halpern 2001; Pavlov et al. 2001) an X-ray pulsar wind nebula (PWN), which has been modelled in detail by Helfand et al. (2001) with an alternative interpretation by Radhakrishnan \& Deshpande (2001).  X-ray and synchrotron emissions are closely linked as highly charged particle flows drive both, so radio emission is a natural comparison.  This targeted observation was optimised at 6~cm to match the Chandra image resolution and to improve on previous radio knowledge.  We imaged a compact radio counterpart about the pulsar (figure 1).  Previous radio studies (Bietenholz, Frail, \& Hankins 1991; Frail et al. 1997a; Bock et al. 1998a; Bock, Turtle, \& Green 1998b) have focussed on larger scales of filaments or wisps and the apparent connection to the Vela-X region of the supernova remnant.  Reprocessed ATCA archive data showed diffuse extended emission in the region of the lobes and indicated a lack of emission in the X-ray PWN region, but it was not clear if this was due to sensitivity limits or a signature of radio emission.

\section{Observation}  The ATCA correlator sampling rate gives a minimum bin size of 2.5~ms over 32~bins, which is within the 89~ms period of the Vela pulsar.  The on-pulse flux density was contained in one bin over all observations due to the precise timing solution from our real-time monitoring project at the University of Tasmania (Dodson \& Lewis, these proceedings) and the small duty cycle (2.5\%).  Binning allows the strong, time-variable pulse flux density to be cleanly removed from the residual image, which has a comparable flux to the mean pulse flux density.  The on-pulse bin flux density is 2.0~Jy, while the combined mean flux density level for off-pulse was 0.1~Jy at 4800~MHz.  The on-pulse bin image gives an excellent phase self-calibrator for the residual image of 31-bins by using the strong point source, ie. the Vela pulsar, which is usually unavailable for otherwise less luminous pulsars.

Previous comparable ATCA observations have only used 8~bins, while the VLA is limited to pulsar `gating' (Frail et al. 1991).  Vela is at low elevations for the VLA, where-as the ATCA can observe Vela for full {\em u-v} coverage at high elevations.  The narrower bin of the 32-binned signal increases the signal-to-noise ratio of the on-pulse bin compared with previous 8-bin observations.  Post-detection de-dispersion is more effective across 32~bins, to a level where smearing is insignificant across the 128~MHz bandwidths.  Improving de-dispersion and the number of off-pulse bins increased the effective off-pulse time from 75\% to 93\%.  Increasing off-pulse information and self-calibration also reduces the time spent on the secondary calibrator (previously up to 13\% plus slew time).

A full complement of ATCA configurations (6D, 1.5D, 0.75D and 0.375D) gave angular scales of 1' to 10" (4 to 20~k$\lambda$). This maximised and included shorter baselines making it more suited to the 30" scale size desired than previous observations.  The longer baselines were included to subtract the relatively strong FR-II background galaxy at 08:35:23,~-45:07:35.  The two 128~MHz IF's were placed in a single band with central frequencies of 4800~MHz and 5696~MHz, increasing {\em u-v} coverage and sensitivity giving a final image RMS of 30~${\mu}$Jy/beam.

\section{Results}  Figure 1 shows a northern and southern lobe with overlayed contours from archive Chandra X-ray data.  The structure is smooth and resolved on the 6~km baselines.  Emission is oriented in projection around the spin axis of the pulsar (Helfand et al. 2001; Radhakrishnan \& Deshpande 2001) and the radio VLBI (Legge 2001) and optical HST (Caraveo et al. 2001) proper motions.  There is no wake of emission behind the proper motion of the pulsar.  The emission is typically 70\% linearly polarised.  In figure 2, the spectral index comparison is between C-band (effective central frequency of 5230~MHz) and archived (Bock et al. 1998) S-band data (central frequency 2368~MHz).  The C-band total intensity flux density contours of 0.5 and 1.0~mJy are overlayed.  Values increase past -1.5 in the southern lobe but are less reliable (nearing the S/N cut-off).  Figure 3 shows an extrapolation of the Radhakrishnan \& Deshpande (2001) work from an X-ray to a radio model of emission from the pulsar.

\section {Discussion}  The morphology is inconsistent with a bowshock PWN, the size is well below that of a static shock PWN and the spin down energy to radio luminosity ratio of $3\times10^{32}$ is much lower than the upper limits found for other radio PWN (Frail \& Scharringhausen 1997; Gaensler et al. 2000).  The velocities given by Pavlov et al. (2001) are sufficient to deliver the particles to the radio nebula, but the lack of a wake and small radius of emission from the pulsar requires it to cease within a few thousand years.  No conventional energy loss mechanism can achieve this; synchrotron spectral aging gives 10$^{5}$ years (Lang 1980), while expansion and inverse Compton processes are also insufficient.  Possibly a reverse shock in the SNR may be suppressing the emission (Blondin, Chevalier, \& Frierson 2001).  There is a bridge of emission from the northern lobe to a peak at 08:35:45~-45:15.  This is consistent with extrapolating an assumed characteristic age of 11 kyr with the proper motion representing the birth site and is within errors of a TeV $\gamma$-ray source (Yoshikoshi et al. 1997), previously noted by Bock et al. (1998).  Proper motion and spin axis, if in agreement, may support theories of slow neutron star kick mechanisms (Lai, Chernoff, \& Cordes 2001).  The three-dimensional orientation can be determined from the radio results, X-ray and X-ray/radio theory (Helfand et al. 2001; Radhakrishnan \& Deshpande 2001).  The radio lobes lie around the spin axis in agreement with models where the emission is driven by particles from the magnetic poles.  The high polarisation and spectral index of -0.5 within the contours (figure 2) implies that the emission is due to a synchrotron source.  The bright northern edge (previously described as a 'wisp') polarisation and spectral index characteristics were separated from Vela-X previously by Beitenholz et al. (1991).  The intensity, polarisation and spectral index characteristics of the lobes are different to the Vela-X region.  We propose that this shows that the compact nebula is directly driven by the pulsar, while the formation of the Vela-X region is a longer term process.

\pagebreak
\begin{figure}
  \psfig{file=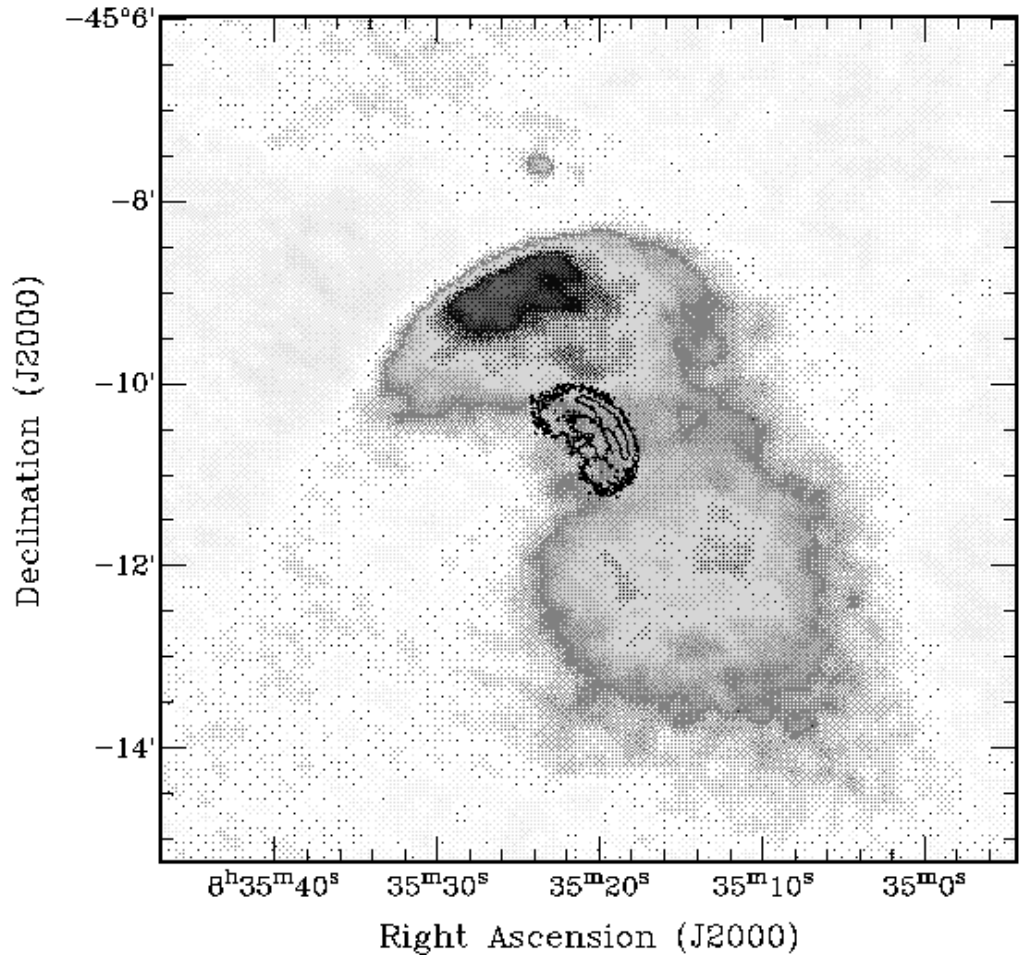,width=13.5cm}
  \caption{6~cm image with X-ray contours} 
  \label{fig:main image}
\end{figure}

\begin{figure}
  \psfig{file=lewisd_2.ps,width=7cm,angle=270}
  \vspace{-5.9cm} \hspace{7.4cm}
  \psfig{file=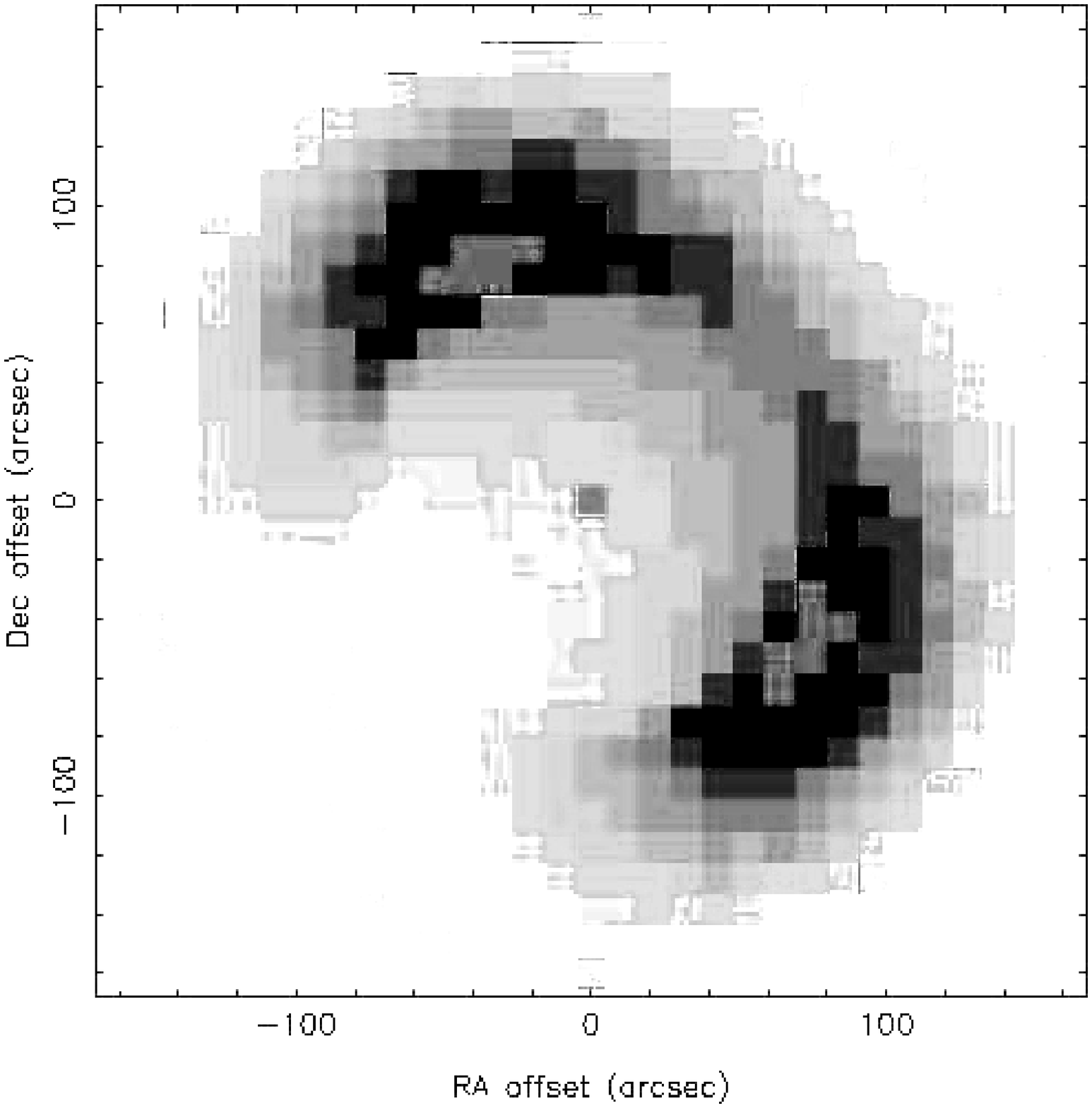,width=5.8cm,angle=270}
  \caption{Spectral index map~~~~~~~~~~~~~~~~Figure 3.  Theoretical model} 
  \label{fig:spectral and theoretical}
\end{figure}

\end{document}